\documentclass[12pt,a4paper]{article}
\usepackage[DIV12]{typearea}
\usepackage{graphicx}
\usepackage{amsfonts}
\usepackage{amsmath}
\usepackage{mathpazo}

\title{The Red Queen visits Minkowski Space}
\author{Robert J Low\thanks{mtx014@coventry.ac.uk}\\
Department of Mathematical Sciences\\
Coventry University\\
Priory Street\\
Coventry CV1 5FB\\
UK}
\date{}
\begin{document}

\maketitle


\begin{abstract}
When Alice went \textit{Through the Looking Glass} \cite{lc}, she 
found herself in a situation where she had to run as fast as she 
could in order to stay still. In accordance with the dictum that 
truth is stranger than fiction, we will see that it is possible to 
find a situation in special relativity where running towards one's 
target is actually counter-productive. Although 
the situation is easily analysed algebraically, the qualitative properties 
of the analysis are greatly illuminated by the use of space-time diagrams.
\end{abstract}

Although tachyons (particles which travel faster than light) are not at 
present observed experimentally, they arise naturally in superstring theory, 
where their consequences require investigation: one example of such an inquiry
is found in \cite{sen}. Outside this context, tachyons have also been
considered from advanced viewpoints, as in \cite{fein}, in which it was
found that the obvious problems associated with causality might be
illusory; and from elementary viewpoints, as in \cite{lb} where
simple geometrical properties of a tachyonic wavefront were considered. 

This article takes a brief look at how tachyons appear to move from the 
point of view of various inertial observers in special relativity. The
results are reminiscent of Alice's experience through the looking glass,
where she had to run as fast as she could just to stay still. Here we
will find that the situation can be worse even than that: it is possible
for a target to recede faster, the faster you chase it.

Although the results are easy to obtain algebraically, it is the use 
of space-time diagrams that renders the situation intelligible. Finally, 
the relative strengths of the algebraic and diagrammatic approaches are 
briefly discussed. The article is presented in a discursive manner, and 
should be accessible to students who have taken a course in special 
relativity.

We will restrict our attention to situations in which all motion takes 
place along the $x$-axis in Minkowski space, so that we can consider 
kinematics in a two-dimensional space-time; furthermore, we will suppose 
that units have been chosen so that $c=1$.  So, consider an inertial 
frame $\Sigma$, with associated coordinates $(t,x)$. In this space-time 
we have three observers, $A$, $B$ and $C$. $A$ is at rest at $x=0$, 
$B$ is heading in the positive $x$-direction with speed $1/2$ and 
$C$ is travelling in the negative $x$-direction  at speed $1/2$; 
they all meet at $x=0$ when $t=0$.

Now consider the following announcements, made just after the 
three observers pass each other:\\
\begin{tabular}{ll}
A & I just saw something travelling in the positive $x$-direction at 
speed $u_A=3/2$\\
B & I just saw something travelling in the positive $x$-direction at 
speed $u_B=4$\\
C & I just saw something travelling in the positive $x$-direction 
at speed $u_C=8/7$
\end{tabular}

The surprising thing is that these three comments should all apply to 
observations of the same object. The reason it is surprising is that 
since $B$ is travelling in the positive $x$-direction, and $C$ in 
the negative $x$-direction, we would normally expect that $B$ should 
see an obect travelling in the positive $x$-direction travel slower 
than $A$, while $C$ would see it travel faster. But, contrariwise, 
the observations have $u_C < u_B < u_A$.

In order to resolve this apparent paradox, let us consider how 
velocities transform between frames of reference in special 
relativity. We will see that this is, in fact, independent
of the velocity to be transformed, and so the usual relativistic
`addition of velocities' is valid even when we are working with
a tachyonic particle.

For simplicity, we consider only one dimension of space. 
So let  $\Sigma$, with coordinates $(t,x)$, be some nominal rest 
frame, and let $\Sigma'$ be a frame whose origin is travelling with 
speed $V$ in the positive $x$-direction in $\Sigma$, with coordinates
$(t',x')$. Suppose also that the event with $t=0,x=0$ also has 
$t'=0,x'=0$. Then the coordinates $(t,x)$ and $(t',x')$ are related 
by the usual Lorentz transformation
\begin{equation}
\label{lt}
\begin{split}
t &= \gamma (t'+Vx') \\
x &= \gamma(x'+Vt')
\end{split}
\end{equation}
where $\gamma = 1/\sqrt{1-V^2}$.

Now, suppose we have an object whose world-line is given in terms of 
$(t,x)$ by $x=ut$, so that it is travelling with speed $u$ in the 
positive $x$-direction. Expressing $x$ and $t$ in terms of $x'$ 
and $t'$ we immediately obtain
\[
x'+Vt' = u(t'+Vx')
\]
which is easily  rearranged to give
\[
x' = \frac{u-V}{1-uV}t'.
\]
Denoting by $u'$ the speed in the positive $x'$-direction, as
measured in $\Sigma'$, we have
\[
u' = \frac{u-V}{1-uV}.
\]
Thus we have the usual `addition of velocities' rule, and 
observe that this result is quite 
independent of the sign or size of $u$. 

So we can now easily check that if $A$ sees an object moving to 
the right at speed $3/2$, then $B$ (for whom $V=1/2$) will attribute 
to it a speed of 
\[
\frac{3/2 - 1/2}{1-3/4} = 4
\]
while $C$ (with $V=-1/2$)  will find its speed to be
\[
\frac{3/2+1/2}{1+3/4} = 8/7.
\]

We can see, then, from the algebraic properties of the Lorentz 
transformations, that this is indeed how the velocity of a 
tachyonic particle would transform between frames of reference. 
In fact, the speed of the tachyon as measured by a moving observer 
has still more peculiar properties.

First, let us look at the tachyon's velocity in a frame moving 
with velocity $V$; if we call this velocity $u'$, then we saw
above that
\[
u' = \frac{u-V}{1-uV}.
\]
Differentiating this with respect to $V$, we obtain
\[
\frac{\partial u'}{\partial V} = \frac{u^2 - 1}{(1-uV)^2}
\]
which is always positive; hence, the faster you chase a 
tachyon, the faster it recedes.

However, even this is not as straighforward as it looks at 
first glance. Examining the form of $u'$ more carefully, 
we make the following observations:
\begin{enumerate}
\item As $V \rightarrow -1$, $u' \rightarrow 1$
\item For $V$ between $-1$ and $1/u$, $u'$ is increasing, 
and $u' \rightarrow \infty$ as $V \rightarrow 1/u$ from below.
\item As $V \rightarrow 1/u$ from above, $u' \rightarrow -\infty$.
\item $u'$ is increasing as $V$ increases from $1/u$ to $1$, 
and as $V \rightarrow 1$, $u' \rightarrow 1$.
\end{enumerate}

So we see that the tachyon is seen to travel faster than light 
by all inertial observers; but that as the speed of the moving 
observer increases, the speed with which the tachyon recedes 
increases without bound until suddenly it switches from receding
with extremely high speed to approaching with extremely high
speed, but then the speed of approach decreases as the
speed of the moving observer continues to increase.

Again, although it is simple 
to derive all this by applying simple algebra to the velocity 
transformation rule, it is unclear what is really
going on here.

In order to obtain some insight into the situation, we 
consider some space-time diagrams \cite{hm}. (It is worth 
noting that this form of space-time diagram is not the 
only one; Shadowitz \cite{sh} considers a variety of 
space-time diagrams, each of which has its strengths. 
However, we will make use of only the form due to Minkowski, 
and leave investigation of the others to the reader.) 

First, consider a space-time diagram that shows only the 
rest-frame of $A$, namely $\Sigma$, and the tachyon worldline 
in it. As is customary, the units of distance and time are 
chosen such that light rays are at $45^\circ$ to the vertical.

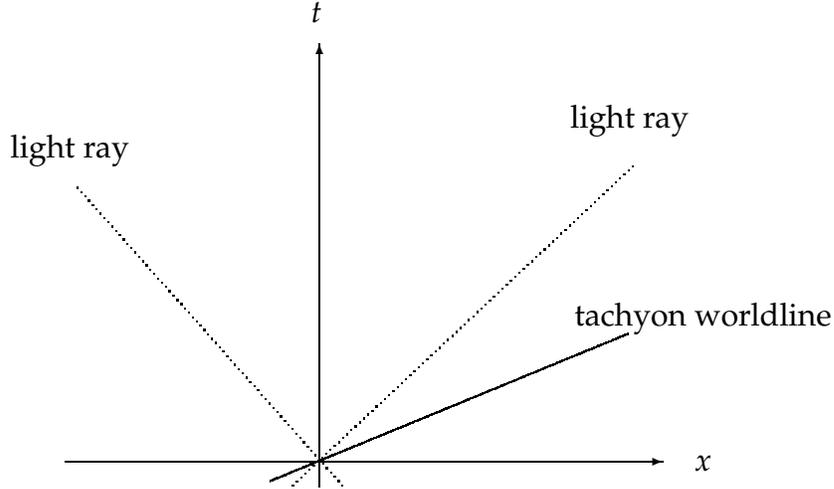
\begin{center}
\unitlength 0.8mm
\linethickness{0.4pt}
\ifx\plotpoint\undefined\newsavebox{\plotpoint}\fi 
\begin{picture}(111.5,82.75)(0,0)
\put(5.5,8){\vector(1,0){99.25}}
\put(47.75,3.75){\vector(0,1){73.75}}
\put(111.5,7.75){\makebox(0,0)[cc]{$x$}}
\put(47.25,82.75){\makebox(0,0)[cc]{$t$}}
\multiput(43.18,3.93)(.69753,.65432){82}{{\rule{.4pt}{.4pt}}}
\multiput(51.43,3.93)(-.63768,.71739){70}{{\rule{.4pt}{.4pt}}}
\multiput(39.5,4.75)(.0818431912,.0337001376){727}{\line(1,0){.0818431912}}
\put(99.25,64.75){\makebox(0,0)[cc]{light ray}}
\put(6.25,59.75){\makebox(0,0)[cc]{light ray}}
\put(111.5,31.75){\makebox(0,0)[cc]{tachyon worldline}}
\end{picture}
\end{center}
So we clearly see that the tachyon is proceeding in the 
direction of increasing $x$ faster than a light 
ray in frame $\Sigma$.

We can now introduce to the diagram the $t'$ and $x'$ axes 
of the rest frame of an observer moving at constant
velocity. First, we note that the relation given in
equation \ref{lt} can be rearranged to give
\[
\begin{split}
t' &= \gamma(t-Vx)\\
x' &= \gamma(x-Vt)
\end{split}
\]
so that the $x'$-axis is given by $t'=0$, i.e. $t=Vx$,
and the $t'$-axis by $x'=0$, i.e. $t=x/V$.

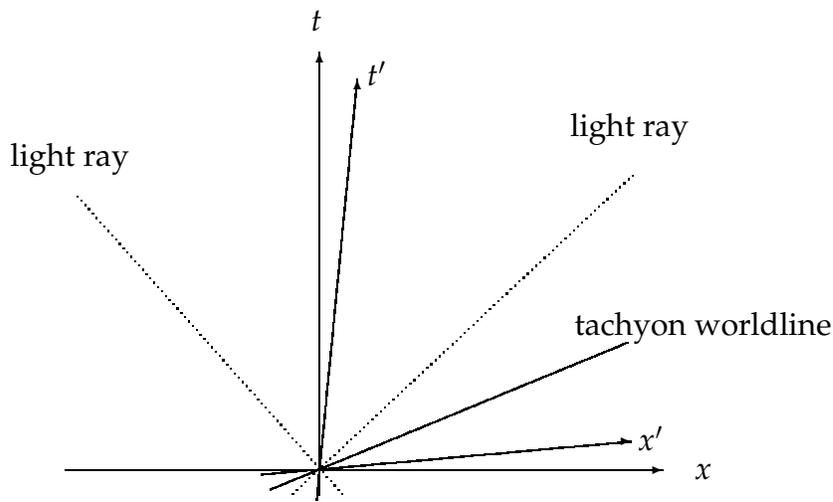
\begin{center}
\unitlength 0.8mm
\linethickness{0.4pt}
\ifx\plotpoint\undefined\newsavebox{\plotpoint}\fi 
\begin{picture}(111.5,82.75)(0,0)
\put(5.5,8){\vector(1,0){99.25}}
\put(47.75,3.75){\vector(0,1){73.75}}
\put(111.5,7.75){\makebox(0,0)[cc]{$x$}}
\put(47.25,82.75){\makebox(0,0)[cc]{$t$}}
\multiput(43.18,3.93)(.69753,.65432){82}{{\rule{.4pt}{.4pt}}}
\multiput(51.43,3.93)(-.63768,.71739){70}{{\rule{.4pt}{.4pt}}}
\multiput(39.5,4.75)(.0818431912,.0337001376){727}{\line(1,0){.0818431912}}
\put(99.25,64.75){\makebox(0,0)[cc]{light ray}}
\put(6.25,59.75){\makebox(0,0)[cc]{light ray}}
\put(111.5,31.75){\makebox(0,0)[cc]{tachyon worldline}}
\put(99.5,12.75){\vector(1,0){.07}}\multiput(38,7.25)(.375,.03353659){164}{\line(1,0){.375}}
\put(54,73){\vector(0,1){.07}}\multiput(47.25,3)(.03358209,.34825871){201}{\line(0,1){.34825871}}
\put(102.75,13.25){\makebox(0,0)[cc]{$x'$}}
\put(57.25,74){\makebox(0,0)[cc]{$t'$}}
\end{picture}
\end{center}

By inspecting this diagram we see that the situation is not 
so counter-intuitive after all. In the same way as a Lorentz 
transformation to a frame with positive speed in the $x$-direction 
will make the worldline of a particle travelling in that direction 
with a lesser speed `more timelike' in the sense that it becomes 
nearer the $t$-axis, such a Lorentz transformation  will make 
the worldline of a tachyon `more spacelike'; so that an observer 
who is travelling in the same direction as a tachyon attributes 
to it a greater speed than the stationary observer. Furthermore, 
as the speed increases, it reaches a value at which the tachyon 
worldline is a line of simultaneity (the perceived speed taking 
on unboundedly large values); and for larger speeds, the 
tachyon worldline has speed with unboundedly large value 
to the left, which then reduce in magnitude as the speed 
of the observer continues to grow.

The diagrammatic investigation also brings out the symmetry 
between this situation and that of the different velocities 
ascribed by observers to an object moving with constant 
sub-luminal speed; for just as by chasing sufficiently 
fast an observer can make this object's worldline will 
pass through the his line of constant position, so he 
can make the tachyon's worldline pass through his line 
of constant time. 

From this investigation, then, we can see the respective strengths 
of the algebraic and diagrammatic approaches to analysing this 
situation. The algebraic approach provides complete quantitative 
information, but does little to give any insight into the 
qualitative behaviour of the transformed velocities. On the 
other hand, space-time diagrams make the qualititative behaviour 
comprehensible, without giving easy access to the numerical 
values of observed speed.

\end{document}